\documentclass{article}
\usepackage{arxiv}

\usepackage[utf8]{inputenc} 
\usepackage[T1]{fontenc}    
\usepackage{hyperref}       
\usepackage{url}            
\usepackage{booktabs}       
\usepackage{amsfonts}       
\usepackage{nicefrac}       
\usepackage{microtype}      
\usepackage{lipsum}		
\usepackage{graphicx}
\usepackage{natbib}
\usepackage{subcaption}
\usepackage{doi}
\usepackage{amsmath,amssymb,amsfonts}%
\usepackage{amsthm}%
\usepackage{mathrsfs}%
\usepackage{bbm}%
\usepackage{textcomp}%
\usepackage{manyfoot}%
\usepackage{bm}%
\usepackage{threeparttable}%
\usepackage{color}%

\title{Stochastic EM Estimation and Inference for Zero-Inflated Beta-Binomial Mixed Models for Longitudinal Count Data}

\date{}

\author{\href{https://orcid.org/0009-0005-5811-7062}{\includegraphics[scale=0.06]{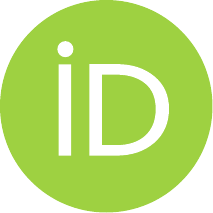}\hspace{1mm}
    John Barrera}
    \thanks{Corresponding author. E-mail: \texttt{john.barrera@postgrado.uv.cl}} \\
	  Instituto de Ingeniería Matemática \\
    Facultad de Ingeniería \\
    Universidad de Valparaíso \\
    Valparaíso, Chile\\
    \And
    \href{https://orcid.org/0000-0002-2398-3156}{\includegraphics[scale=0.06]{pix/orcid.pdf}\hspace{1mm}
    Ana Arribas-Gil} \\
	Departamento de Estadística \\ 
    Universidad Carlos III de Madrid \\
    Getafe, Spain\\
    \And
    \href{https://orcid.org/0000-0002-8995-8535}{\includegraphics[scale=0.06]{pix/orcid.pdf}\hspace{1mm}
    Dae-Jin Lee} \\
	School of Science and Technology \\ 
    IE University \\
    Madrid, Spain\\
	\And
	\href{https://orcid.org/0000-0001-6332-6137}{\includegraphics[scale=0.06]{pix/orcid.pdf}\hspace{1mm}
    Cristian Meza} \\
	CIMFAV \\
    Universidad de Valparaíso \\
    Valparaíso, Chile	
}

\renewcommand{\headeright}{}
\renewcommand{\undertitle}{}
\renewcommand{\shorttitle}{Stochastic estimation of a zero-inflated beta-binomial mixed model}

\hypersetup{
pdfauthor={J. Barrera et al.}
}

\begin{document}
\maketitle

\begin{abstract}
	Analyzing overdispersed, zero-inflated, longitudinal count data poses significant modeling and computational challenges, which standard count models (e.g., Poisson or negative binomial mixed effects models) fail to adequately address. We propose a Zero-Inflated Beta-Binomial Mixed Effects Regression (ZIBBMR) model that augments a beta-binomial count model with a zero-inflation component, fixed effects for covariates, and subject-specific random effects, accommodating excessive zeros, overdispersion, and within-subject correlation. Maximum likelihood estimation is performed via a Stochastic Approximation EM (SAEM) algorithm with latent variable augmentation, which circumvents the model's intractable likelihood and enables efficient computation. Simulation studies show that ZIBBMR achieves accuracy comparable to leading mixed-model approaches in the literature and surpasses simpler zero-inflated count formulations, particularly in small-sample scenarios. As a case study, we analyze longitudinal microbiome data, comparing ZIBBMR with an external Zero-Inflated Beta Regression (ZIBR) benchmark; the results indicate that applying both count- and proportion-based models in parallel can enhance inference robustness when both data types are available.
\end{abstract}

\keywords{zero-inflated beta-binomial, mixed-effects, SAEM, microbiome, longitudinal data, overdispersion}

\section{Introduction}
\label{sec:intro}

Longitudinal count data are common in biomedical, environmental, and social sciences, and often exhibit three key complexities: excess of zeros, overdispersion, and within-subject correlation due to repeated measurements. Examples include microbial sequence counts in microbiome studies \citep{Romero2014}, parasite loads in epidemiological surveys, and defect counts in manufacturing processes. Standard count models, such as Poisson or negative binomial mixed-effects models, typically cannot address all these features simultaneously, which may lead to biased estimates, unreliable inference, and reduced predictive performance \citep{mccullagh1989generalized, hall2005zero}.  

The beta-binomial distribution is well suited for modeling overdispersed binomial data \citep{Skellam1948,Najera2019}, and zero-inflated models can capture structural zeros beyond those expected under the assumed count distribution \citep{lambert1992zero}. When repeated measurements are taken, mixed-effects structures can account for unobserved heterogeneity among subjects \citep{pinheiro2006mixed}. However, combining zero inflation, beta-binomial modeling, and subject-specific random effects produces a likelihood function that is analytically intractable, making maximum likelihood estimation nontrivial.  

We address this challenge by introducing the \emph{Zero-Inflated Beta-Binomial Mixed Effects Regression} (ZIBBMR) model that simultaneously accounts for structural zeros, extra-binomial variation, and within-subject correlation in longitudinal binomial-type data. Parameters are estimated via a maximum likelihood estimation procedure based on the stochastic approximation expectation--maximization (SAEM) algorithm \citep{delyon1999convergence}. This approach incorporates latent variable augmentation and stochastic simulation to overcome the computational difficulties, enabling likelihood-based inference for complex zero-inflated mixed models. In particular, the SAEM algorithm is an alternative to the EM algorithm when the Expectation step is intractable, which is overcome by approximating the conditional expectation through simulation while preserving the convergence properties of the EM algorithm.
The remainder of the paper is organized as follows. Section~2 presents the Zero-Inflated Beta–Binomial Mixed Effects Regression (ZIBBMR) model, its hierarchical representation, and the maximum likelihood estimation via the SAEM algorithm, including the hierarchical augmentation and implementation details. Section~3 reports simulation studies that compare the proposed estimator against general estimation approaches for generalized linear mixed models as \texttt{glmmTMB} \citep{Brooks2017,McGilly2025} or generalized additive models as \texttt{gamlss} \citep{Rigby2005,Stasinopoulos2017}, describing the design, performance metrics, and results. Section~4 turns to the application, providing background and motivation for a longitudinal vaginal microbiome dataset and comparing ZIBBMR with an alternative Zero–Inflated Beta Regression (ZIBR) benchmark. Section~5 concludes with a discussion and future directions. An \textsf{R} implementation and reproducible materials are summarized in the \textit{Software} section and the supplementary material.

\section{Model definition, estimation and inference}\label{sec:model}

\subsection{Definition of the Zero-Inflated Beta-Binomial Mixed Regression (ZIBBMR) model}\label{ssec:def}

Let $Y_{it}$ be the observed count for subject $i$ ($1 \leq i \leq N$) at measurement occasion $t$ ($1 \leq t \leq T_i$), with $S_{it}$ the total number of trials, which is known.  
For convenience, denote the full set of observations by
$$
\mathbf{Y} = \left\{ Y_{it} : 1 \leq i \leq N, \; 1 \leq t \leq T_i \right\},
$$
which represents all responses across subjects and measurement occasions.  

Conditional on covariates and subject-specific random effects, the Zero-Inflated Beta-Binomial Mixed Regression (ZIBBMR) model assumes
\begin{equation}\label{equ.zibb1}
    Y_{it} \sim
    \begin{cases}
        0, & \text{with probability } 1-p_{it},\\
        \operatorname{BetaBin}\!\left(S_{it}, u_{it}\phi,(1-u_{it})\phi\right), & \text{with probability } p_{it},
    \end{cases}
\end{equation}
where $\phi>0$ and $0 < u_{it}, p_{it} < 1$. The parameters $p_{it}$ (zero-inflation probability) and $u_{it}$ (mean success probability) are linked to covariates via
\begin{equation}\label{equ.zibb2}
    \mathrm{logit}(p_{it}) = a_i + X_{it}^\top\alpha, \quad
    \mathrm{logit}(u_{it}) = b_i + Z_{it}^\top\beta,
\end{equation}
where $X_{it}$ and $Z_{it}$ are covariate vectors, $\alpha$ and $\beta$ are regression coefficient vectors, and $(a_i, b_i)^\top$ are independent random intercepts with $a_i \sim N(a, \sigma_1^2)$ and $b_i \sim N(b, \sigma_2^2)$.

The Beta-Binomial probability mass function is given by
\begin{equation}\label{equ.bbdens}
\begin{split}
P\left(Y = Y_{it}|S_{it}, u_{it}, \phi\right)
&= f\left(Y_{it};S_{it}, u_{it}, \phi\right) \\
&= {S_{it} \choose Y_{it}}\frac{B\left(Y_{it} + u_{it} \phi,S_{it} - Y_{it} + (1 - u_{it}) \phi\right)}{B\left(u_{it} \phi,(1 - u_{it}) \phi\right)},
\end{split}
\end{equation}
where $B(\cdot,\cdot)$ denotes the beta function.  

From Eq.~\eqref{equ.bbdens}, the mean and variance are given by
\begin{equation}\label{equ.bbmean}
E\!\left(Y_{it} \,\middle|\, S_{it}, u_{it}, \phi\right) 
= S_{it} u_{it},
\end{equation}
and
\begin{equation}\label{equ.bbvar}
\mathrm{Var}\!\left(Y_{it} \,\middle|\, S_{it}, u_{it}, \phi\right) 
= S_{it} u_{it} (1 - u_{it}) 
  \left[ 1 + \frac{S_{it} - 1}{\phi + 1} \right].
\end{equation}

The variance expression in Eq.~\eqref{equ.bbvar} contains an additional term compared to the classical binomial variance, allowing the Beta-Binomial distribution to capture overdispersion in discrete data \citep{Skellam1948}.

If we consider the following parametrization for the density function of the Beta distribution
\begin{equation}\label{equ.betadens}
f_{\mathrm{Beta}}(x; \alpha, \beta)
= \frac{x^{\alpha - 1} (1 - x)^{\beta - 1}}{B(\alpha, \beta)},
\quad 0 < x < 1,
\end{equation}
then, in the ZIBBMR parameterization,
$\alpha = u_{it} \phi$ and $\beta = (1-u_{it}) \phi$.

An equivalent hierarchical representation is
\begin{align}\label{equ.zibbjer}
    Y_{it} \mid S_{it}, p_{it}, w_{it}, \phi &\sim 
    \begin{cases}
        0, & \text{with probability } 1-p_{it},\\
        \operatorname{Bin}(S_{it}, w_{it}), & \text{with probability } p_{it},
    \end{cases}\\
    w_{it} \mid u_{it}, \phi &\sim \operatorname{Beta}(u_{it}\phi,(1-u_{it})\phi), \nonumber
\end{align}
with $w_{it}$ an unobserved latent variable. This formulation expresses the Beta-Binomial as a Binomial distribution with a Beta-distributed success probability, and facilitates a data augmentation approach for inference.

From \eqref{equ.zibb1} and \eqref{equ.zibb2}, the observed-data likelihood for $\bm\theta=(\phi,a,b,\alpha,\beta,\sigma_1^2,\sigma_2^2)$ is
\begin{equation}\label{equ.zibblik}
\begin{split}
L(\theta; \mathbf{Y}) 
&= \prod_{i=1}^N \int_{\mathbb{R}} \! \int_{\mathbb{R}}
    \prod_{t=1}^{T_i} 
    (1 - p_{it})^{\mathbbm{1}_{\{Y_{it} = 0\}}} \\
&\quad \times 
    \left[ p_{it} \, f\!\left(Y_{it}; S_{it}, u_{it}, \phi\right) \right]^{\mathbbm{1}_{\{Y_{it} > 0\}}}
    g(a_i,b_i) \, da_i \, db_i .
\end{split}
\end{equation}
where $g(\cdot,\cdot)$ is the joint density of the normal random effects. This integral has no closed form, motivating a stochastic approximation approach for maximum likelihood estimation.

\subsection{Estimation via the Stochastic Approximation EM Algorithm}\label{ssec:estim}

The Stochastic Approximation Expectation-Maximization (SAEM) algorithm \citep{delyon1999convergence} is an iterative procedure for Maximum Likelihood estimation in latent variable models in which the Expectation step of the EM algorithm is not explicit. Let $\mathbf{Y}=\left(Y_{it}, 1\leq i\leq N, 1\leq t\leq T_i\right)$ and $\boldsymbol{\varphi}=\left(\varphi_i, 1\leq i\leq N\right)$ denote the observed and non-observed data, respectively, so the complete data of the model are $(\mathbf{Y},\boldsymbol{\varphi})$. In this case, the SAEM algorithm consists of replacing the usual E-step of EM with a stochastic approximation procedure. Given an initial point $\bm\theta^{(0)}$, iteration $q$ of the algorithm writes:
\begin{itemize}
    \item \textbf{Simulation (S) step:} Draw a realization $\boldsymbol{\varphi}^{(q)}$ from the conditional distribution $p\left(\cdot|\;\mathbf{Y};\bm\theta^{(q-1)}\right)$.
    \item \textbf{Stochastic Approximation (SA) step:} Update $s_{q}(\bm\theta)$, the approximation of the conditional expectation $\mathbb{E}\left[\log p\left(\mathbf{Y},\boldsymbol{\varphi}^{(q)};\bm\theta\right)|\mathbf{Y},\bm\theta^{(q-1)}\right]$:
    $$s_q(\bm\theta)=s_{q-1}(\bm\theta) + \gamma_q \left(\log p\left(\mathbf{Y}, \boldsymbol{\varphi}^{(q)}; \bm\theta\right)- s_{q-1}(\bm\theta)\right)$$
\noindent where $\{\gamma_q\}_{q\in\mathbb{N}}$ is a decreasing sequence of stepsizes with $\gamma_1=1$. 
    \item \textbf{Maximization (M) step:} Update $\bm\theta^{(q)}$ according to $\bm\theta^{(q)}= \arg\max_{\bm\theta} s_q(\bm\theta).$
\end{itemize}

In the ZIBBMR model let $\varphi_i = (a_i, b_i)$, $1 \le i \le N$, denote the unobserved random effects, with $\boldsymbol{\varphi} = (\varphi_i)_{i=1}^N$.  
By construction, $\varphi_i \sim N(\boldsymbol{\mu}, \mathbf{G})$, with $\boldsymbol{\mu} = (a, b)$ and $\mathbf{G} = \mathrm{diag}(\sigma_1^2, \sigma_2^2)$.

The complete-data likelihood factorizes as
\begin{equation}\label{equ.decomp3}
\begin{split}
p(\mathbf{Y}, \boldsymbol{\varphi}; \bm\theta)
&= p(\mathbf{Y} \mid \boldsymbol{\varphi}; \alpha, \beta, \phi) \;
   p(\boldsymbol{\varphi} \mid \boldsymbol{\mu}, \mathbf{G}) \\
&\propto |\mathbf{G}|^{-N/2} \prod_{i=1}^N 
   \exp\!\left\{ -\frac{1}{2} (\varphi_i - \boldsymbol{\mu})^\top \mathbf{G}^{-1} (\varphi_i - \boldsymbol{\mu}) \right\} \\
&\quad \times \prod_{i=1}^N \prod_{t=1}^{T_i}
    (1 - p_{it})^{\mathbbm{1}_{\{Y_{it} = 0\}}} \;
    p_{it}^{\mathbbm{1}_{\{Y_{it} > 0\}}} \;
    f\!\left(Y_{it}; S_{it}, u_{it}, \phi\right)^{\mathbbm{1}_{\{Y_{it} \ge 0\}}}.
\end{split}
\end{equation}

\paragraph{SAEM update scheme}
Given a starting point $\bm\theta^{(0)}$ and iteration index $q \ge 1$, the SAEM algorithm for ZIBBMR proceeds as follows:
\begin{enumerate}
\item \textbf{Simulation step:} Draw $\varphi_i^{(q)}$, $i = 1,\ldots,N$, from the conditional distribution $p(\cdot \mid \mathbf{Y}; \bm\theta^{(q-1)})$ (approximated via Metropolis--Hastings; see below).
\item \textbf{Stochastic approximation step:} Update the sufficient statistics  
\begin{equation}\label{equ.update3}
\begin{split}
F_1^{(q)} &= F_1^{(q-1)} + \gamma_q \left( \sum_{i=1}^N \varphi_i^{(q)} - F_1^{(q-1)} \right), \\
F_2^{(q)} &= F_2^{(q-1)} + \gamma_q \left( \sum_{i=1}^N \varphi_i^{(q)} {\varphi_i^{(q)}}^\top - F_2^{(q-1)} \right),
\end{split}
\end{equation}
where $\{\gamma_q\}$ is a decreasing step-size sequence with $\gamma_1 = 1$.
\item \textbf{Maximization step:} Update  
\begin{equation}\label{equ.maxim3}
\begin{split}
\boldsymbol{\mu}^{(q)} &= \frac{F_1^{(q)}}{N}, \\
\mathbf{G}^{(q)} &= \frac{F_2^{(q)}}{N} - \frac{F_1^{(q)} {F_1^{(q)}}^\top}{N^2}.
\end{split}
\end{equation}
\end{enumerate}

Because of the Beta--Binomial component, $(\beta, \phi)$ and $\alpha$ are updated separately:
\begin{align}
\left(\tilde{\beta}^{(q)}, \tilde{\phi}^{(q)}\right)
&= \arg\max_{\beta, \phi} \sum_{i,t} \mathbbm{1}_{\{Y_{it} > 0\}}
   \log\!\left[
      \frac{B\left(Y_{it} + u_{it}^{(q)} \phi, \; S_{it} - Y_{it} + (1 - u_{it}^{(q)}) \phi\right)}
           {B\left(u_{it}^{(q)} \phi, \; (1 - u_{it}^{(q)}) \phi\right)}
   \right], \label{equ.update13} \\
\tilde{\alpha}^{(q)}
&= \arg\max_{\alpha} \sum_{i,t} \left[
   \mathbbm{1}_{\{Y_{it} > 0\}} \log p_{it}^{(q)} +
   \mathbbm{1}_{\{Y_{it} = 0\}} \log(1 - p_{it}^{(q)})
\right], \label{equ.update23}
\end{align}
where $u_{it}^{(q)} = u_{it}(b_i^{(q)}, \beta)$ and $p_{it}^{(q)} = p_{it}(a_i^{(q)}, \alpha)$ via Eq.~\eqref{equ.zibb2}.  
Finally,
\begin{equation}\label{equ.maxim13}
\begin{split}
\phi^{(q)}   &= \phi^{(q-1)} + \gamma_q \left( \tilde{\phi}^{(q)}   - \phi^{(q-1)} \right), \\
\alpha^{(q)} &= \alpha^{(q-1)} + \gamma_q \left( \tilde{\alpha}^{(q)} - \alpha^{(q-1)} \right), \\
\beta^{(q)}  &= \beta^{(q-1)}  + \gamma_q \left( \tilde{\beta}^{(q)}  - \beta^{(q-1)}  \right).
\end{split}
\end{equation}

It is worth noting that the simulation step of the SAEM algorithm can be implemented using a Metropolis-within-Gibbs scheme, as described in Section A of the Supplementary Material. However, this approach is computationally intensive and therefore not well suited to microbiota data, given the typically large size of such data sets. Accordingly, results are reported using a Metropolis–Hastings algorithm. The following section describes the implementation of the Simulation step of the SAEM algorithm.

\subsubsection{Implementation details}\label{ssec:impl}

In practice, the simulation step of SAEM requires careful tuning to ensure efficient exploration of the augmented parameter space and stable convergence.  When using the original specification without augmentation, the random effects $\boldsymbol{\varphi}$ are updated via Metropolis--Hastings, employing three 
proposal kernels:  
(i) a prior proposal $N(\boldsymbol{\mu}^{(q)}, \mathbf{G}^{(q)})$ for global exploration,  
(ii) a multivariate random walk $N(\varphi_{i,p-1}, \Omega^{(q)})$ for correlated local moves,  
and (iii) a univariate random walk on a single randomly chosen component of $\varphi_{i,p-1}$ for fine adjustments.  
In practice, $m_1$ iterations are run with kernel~($kern_1$), $m_2$ with kernel~($kern_2$), and $m_3$ with kernel~($kern_3$) at each SAEM iteration.  
This combination of global and local moves allows the Markov chain to visit widely separated regions of the posterior distribution while also refining estimates in high-probability areas.

The convergence of the algorithm may be enhanced by simulating multiple realizations or independent Markov chains at each iteration and by resorting to a Monte Carlo approximation. More precisely, during the Simulation step, $m$ samples $\varphi_i^{(q,l)} \sim p(\cdot \mid \mathbf{Y}; \bm\theta^{(q-1)})$, for $1 \leq l \leq m$, are generated and in the Stochastic Approximation step, the sufficient statistics are updated according to
\begin{equation}\label{equ.update4}
\begin{split}
F_1^{(q)} &= F_1^{(q-1)} + \gamma_q \left( \sum_{l=1}^m\sum_{i=1}^N \varphi_i^{(q,l)} - F_1^{(q-1)} \right), \\
F_2^{(q)} &= F_2^{(q-1)} + \gamma_q \left( \sum_{l=1}^m\sum_{i=1}^N \varphi_i^{(q,l)} {\varphi_i^{(q,l)}}^\top - F_2^{(q-1)} \right),
\end{split}
\end{equation}

One of the advantages of using multiple sequences or chains is that they can be used to provide empirical estimates of $\mathbb{E}\left(\varphi_i | Y_i ; \bm\theta\right)$ and $\mbox{Var}\left(\varphi_i|Y_i; \bm\theta\right)$. These quantities are required to approximate the observed log-likelihood via Importance Sampling, which in turn allows the computation of the Likelihood Ratio Test (LRT), as described in the next subsection.

\subsubsection{Approximation of the log-likelihood using Importance Sampling}\label{ssec:llis}

For complex mixed effects models, the log-likelihood of the observed data generally does not admit a closed-form expression. Nevertheless, its evaluation is essential for conducting LRTs and for computing information criteria associated with a given model. One possible approximation strategy relies on the Importance Sampling technique \citep{Kloek:1978}. 

Let $\mathcal{LL}_Y(\hat{\bm\theta})$ denote the log-likelihood evaluated at the estimated population parameters, namely $\mathcal{LL}_{\bm{Y}}(\hat{\bm\theta}) = \log p(\bm{Y};\hat{\bm\theta})$, where $p(\bm{Y};\hat{\bm\theta}) = L(\hat{\bm\theta}; \bm{Y})$ represents the joint density of the observed data conditional on $\hat{\bm\theta}$. Note that
$$
\mathcal{LL}_{\bm{Y}}(\hat{\bm\theta}) = \sum_{i=1}^N \log p(Y_i;\hat{\bm\theta}),
$$
and that, for a given \textit{proposal distribution} $\tilde{p}_{\varphi_i}$ absolutely continuous with respect to $p_{\varphi_i}$, the marginal density of $Y_i$ can be written as
$$
p(Y_i;\hat{\bm\theta}) = \int p(Y_i,\varphi_i;\hat{\bm\theta}) \, d\varphi_i = \int p(Y_i|\varphi_i;\hat{\bm\theta}) \frac{p(\varphi_i;\hat{\bm\theta})}{\tilde{p}(\varphi_i;\hat{\bm\theta})}\, \tilde{p}(\varphi_i;\hat{\bm\theta}) \, d\varphi_i.
$$
Hence,
$$
p(Y_i;\hat{\bm\theta}) = \mathbb{E}_{\tilde{p}}\left[p(Y_i|\varphi_i;\hat{\bm\theta}) \frac{p(\varphi_i;\hat{\bm\theta})}{\tilde{p}(\varphi_i;\hat{\bm\theta})}\right].
$$
This expectation can be approximated numerically by:
\begin{enumerate}
    \item Drawing a sample $\varphi_i^{(1)},\ldots,\bm\varphi_i^{(K)}$ from the proposal distribution $\tilde{p}_{\varphi_i}$;
    \item Computing the Monte Carlo average
    $$\hat{p}_{(i,K)} = \frac{1}{K} \sum_{k=1}^K p(Y_i|\varphi_i^{(k)};\hat{\bm\theta})\frac{p(\varphi_i^{(k)};\hat{\bm\theta})}{\tilde{p}(\varphi_i^{(k)};\hat{\bm\theta})}.
    $$
\end{enumerate}

The optimal choice for the proposal distribution would be the conditional density $p_{\varphi_i|Y_i}$, as it yields a zero-variance estimator of the expectation. However, since this conditional distribution is not available in closed form, we instead adopt a proposal distribution that closely approximates it. This proposal is constructed using the empirically estimated conditional mean and variance of the random effects obtained during the simulation step of the SAEM algorithm, namely $\mu_i = \hat{\mathbb{E}}\left(\varphi_i|Y_i ; \hat{\bm\theta}\right)$ and $\sigma_i^2 = \hat{\mathrm{Var}}\left(\varphi_i|Y_i ; \hat{\bm\theta}\right)$. The samples $\varphi_i^{(k)}$, $k=1,\ldots,K$, are then generated according to a noncentral Student $t$ distribution of the form
$$\varphi_i^{(k)} = \mu_i + \sigma_i \, T_{i,k},$$
where $T_{i,k} \stackrel{\text{i.i.d.}}{\sim} t_{\nu}$. Unless stated otherwise, the log-likelihood approximation in this work is computed using $\nu = 5$ and $K = 500$.

\subsubsection{Stochastic approximation of the standard errors}\label{ssec:stder}

Beyond point estimation, it is often desirable for an estimation procedure to provide standard errors of the model parameters. Under maximum likelihood estimation, these can be obtained asymptotically from the Fisher information matrix, which is not analytically available for complex models. Using Louis's missing information principle \citep{Louis1982}, the following identity holds:
$$\partial_{\bm\theta}^2 \log p(\bm{y};\bm\theta)=\mathbb{E}\left(\partial_{\bm\theta}^2 \log p(\bm{Y},\bm\varphi;\bm\theta)|\bm{Y};\bm\theta
\right)+\mathrm{Cov}\left(\partial_{\bm\theta} \log p(\bm{Y},\bm\varphi;\bm\theta)|\bm{y};\bm\theta\right),$$
where
\begin{align*}
  \mbox{Cov}\left(\partial_{\bm\theta}\log p(\bm{Y},\bm{\varphi};\bm\theta)|\bm{Y};\bm\theta\right) =&\mathbb{E}\left(\partial_{\bm\theta}\log p(\bm{Y},\bm{\varphi};\bm\theta)\partial_{\bm\theta}\log p(\bm{Y},\bm{\varphi};\bm\theta)^{\top}|\bm{Y};\bm\theta\right)\\
  &-\mathbb{E}\left(\partial_{\bm\theta}\log p(\bm{Y},\bm{\varphi};\bm\theta)|\bm{Y};\bm\theta\right)\mathbb{E}\left(\partial_{\bm\theta}\log p(\bm{Y},\bm{\varphi};\bm\theta)|\bm{Y};\bm\theta\right)^{\top} 
\end{align*}

Accordingly, $\partial_{\bm\theta}^2 \log p(\bm{y};\bm\theta)$ can be approximated by a stochastic sequence $\{H_q\}_{q\in\mathbb{N}}$ updated as
\begin{eqnarray*}
    D_q &=& D_{q-1} + \gamma_q \left(\partial_{\bm\theta} \log p(\bm{Y},\bm\varphi^{(q)};\bm\theta^{(q)})-D_{q-1}\right),\\
    G_q &=& G_{q-1} + \gamma_q \left(\partial_{\bm\theta}^2 \log p(\bm{y},\bm\varphi^{(q)};\bm\theta^{(q)})\right.\\
    && \left. + \partial_{\bm\theta} \log p(\bm{y},\bm\varphi^{(q)};\bm\theta^{(q)}) \partial_{\bm\theta} \log p(\bm{y},\bm\varphi^{(q)};\bm\theta^{(q)})^{\top} - G_{q-1}\right),
\end{eqnarray*}
with $H_q = G_q - D_q D_q^{\top}$. At convergence, $-H_q^{-1}$ provides an approximation of the covariance matrix of the parameter estimates \citep{Zhu2002,Li2010}, which can subsequently be used to derive hypothesis testing procedures for the model parameters.




\section{Simulation studies}\label{sec:simulation}

This section evaluates the finite-sample performance of the proposed
ZIBBMR estimator based on SAEM, comparing it to two widely used
alternatives: \texttt{glmmTMB} and \texttt{gamlss}. We assess (i) point
estimation accuracy, (ii) coverage of Wald confidence intervals, (iii)
type~I error control, and (iv) computational efficiency, across a range
of sample sizes, zero-inflation levels, and dispersion values.

\subsection{Alternative estimation approaches}\label{sec:sim_alternatives}

While our primary focus is on the SAEM procedure described in
Section~\ref{sec:model}, other estimation strategies are available for
zero-inflated Beta--Binomial regression with random effects:

\begin{itemize}
    \item \textbf{\texttt{glmmTMB}} \citep{Brooks2017,McGilly2025} implements maximum likelihood
    estimation for generalized linear mixed models via Template Model Builder (TMB),
    using Laplace approximation to integrate over random effects. It supports the
    Beta--Binomial distribution and zero-inflation terms in separate linear predictors.
    The optimizer is typically fast and stable, even with moderately complex
    random-effects structures.

    \item \textbf{\texttt{gamlss}} \citep{Rigby2005,Stasinopoulos2017} fits generalized
    additive models for location, scale, and shape by penalized likelihood, allowing
    each distributional parameter to depend on covariates. It supports a
    zero-inflated Beta--Binomial family and can include random effects through
    penalized quasi-likelihood. The method is flexible but penalization can affect
    likelihood ratio test validity \citep{Breslow1993}.
\end{itemize}

In the simulations, all methods are fitted to the same datasets, with identical
covariates, random-effects structures, and distributional assumptions.

\subsection{Design}\label{sec:sim_design}

We consider $N \in \{30,50,100\}$ subjects with $T_i = T \in \{5,10,15\}$
repeated measurements, unless otherwise stated. For subject $i$ and time $t$,
covariates $X_{it}$ and $Z_{it}$ are generated and
\[
\mathrm{logit}(p_{it}) = a_i + X_{it}^\top \alpha^\star,\qquad
\mathrm{logit}(u_{it}) = b_i + Z_{it}^\top \beta^\star,
\]
where $(a_i,b_i)^\top \sim N(\mu^\star, G^\star)$ with
$\mu^\star=(a^\star,b^\star)$ and $G^\star=\mathrm{diag}(\sigma_{1}^{\star 2},\sigma_{2}^{\star 2})$.
Conditional on $(p_{it},u_{it})$,
\[
Y_{it} \sim 
\begin{cases}
0, & \text{with probability } 1-p_{it},\\
\mathrm{BetaBin}\!\big(S_{it},\, u_{it}\phi^\star,\,(1-u_{it})\phi^\star\big), & \text{with probability } p_{it},
\end{cases}
\]
with $S_{it} \stackrel{\mathrm{i.i.d.}}{\sim} \mathrm{Unif}\{200,\dots,800\}$.

\subsection{Data-generating mechanisms}\label{sec:sim_dgp}

We examine four main settings:
\begin{enumerate}
  \item \textbf{Setting 1:} \textit{(same-sign effects, low variance)} with $a = b = -0.5$, $\alpha = \beta = 0.5$, $\sigma_1=0.7$, $\sigma_2=0.5$, $\phi=6.4$.
  \item \textbf{Setting 2:} \textit{(opposite-sign effects, high variance)} with $a=-0.5$, $b=0.5$, $\alpha=0.5$, $\beta=-0.5$, $\sigma_1=1.4$, $\sigma_2=0.8$, $\phi=10.4$.
  \item \textbf{Setting 3:} \textit{(inclusion of covariates, high proportion of zeros)} with $a=-1.8$, $b=-0.9$, $\alpha=(0.8,-0.7)$, $\beta=(0.6,-0.9)$, $\sigma_1=1.35$, $\sigma_2=1.28$, $\phi=12.3$; two covariates $X_1=Z_1$ (binary, 0/1 by subject half) and $X_2=Z_2 \sim N(2,1)$.
  \item \textbf{Setting 4:} \textit{(type I error calibration)} with $a=0.5$, $b=-0.5$, $\alpha=(0,-0.5)$, $\beta=(0,0.5)$, $\sigma_1=0.7$, $\sigma_2=0.5$, $\phi \sim \mathrm{Unif}(2,10)$; two covariates $X_1=Z_1$ (binary, 0/1 by subject half) and $X_2=Z_2 \sim N(1,1)$.
\end{enumerate}

Settings 1--2 use a single binary covariate $X=Z$ (0 for the first $N/2$ subjects, 1 otherwise). Setting 3 mantains the binary covariate and adds a continuous one to test the estimation in a more complete scenario. We also use this Setting to check the calculations of the log-likelihood proposed by the two comparable methods (ZIBBMR-SAEM and \texttt{glmmTMB}). Setting 4 is used to test $H_0:\alpha_1=0$, $H_0:\beta_1=0$, and
$H_0:\alpha_1=\beta_1=0$.

\subsection{Estimators compared}\label{sec:sim_estimators}

We evaluate and compare the following estimation approaches:

\begin{itemize}
  \item \textbf{Proposed:} ZIBBMR--SAEM, as described in Section \ref{ssec:estim}.
  \item \textbf{\texttt{glmmTMB}:} Implementation of the zero-inflated beta-binomial mixed-effects model using the \texttt{glmmTMB} package \citep{Brooks2017}, which relies on Laplace approximation for the integration over random effects.
  \item \textbf{\texttt{gamlss}:} Implementation using the \texttt{gamlss} framework \citep{Rigby2005,Stasinopoulos2017}, which supports beta-binomial distributions with zero inflation via penalized maximum likelihood estimation.
\end{itemize}

\subsection{Performance metrics}\label{sec:sim_metrics}

For each parameter $\theta_j$, performance is evaluated over $R=1000$ Monte Carlo replications using the following metrics:
\begin{itemize}
  \item \textbf{Bias:}
  $$
  \mathrm{Bias} = \frac{1}{R} \sum_{r=1}^R \left( \hat{\theta}_j^{(r)} - \theta_j^\star \right)
  $$
  
  \item \textbf{Root Mean Square Error (RMSE):}
  $$
  \mathrm{RMSE} = \left[ \frac{1}{R} \sum_{r=1}^R \left( \hat{\theta}_j^{(r)} - \theta_j^\star \right)^2 \right]^{1/2}
  $$
  
  \item \textbf{Mean Absolute Error (MAE):}
  $$
  \mathrm{MAE} = \frac{1}{R} \sum_{r=1}^R \left| \hat{\theta}_j^{(r)} - \theta_j^\star \right|
  $$
\end{itemize}

In addition, we compute the p-value of the Wald test for the estimators, as well as the the average computation time per model fit.

\subsection{Implementation details}\label{sec:sim_impl}  

SAEM runs are performed using 5 parallel chains and 1000 iterations for all designs, except for Setting~4, which uses 10 chains to ensure stable estimation of Type~I error rates.  

The initial parameter vectors are selected to be reasonably close to the true values while allowing for sufficient exploration of the parameter space:  
\begin{itemize}
    \item \textit{Settings~1--2:} $\theta^{(0)}=(18,\,-0.3,\,0.2,\,0.8,\,0.1,\,0.48,\,0.72)$  
    \item \textit{Setting~3--4}: $\theta^{(0)}=(6,\,0.4,\,-0.7,\,0.3,\,-0.2,\,0.2,\,0.1,\,0.28,\,0.61)$  
\end{itemize}

These correspond to $(\phi, a, b, \alpha, \beta, \sigma_1, \sigma_2)$. The additional terms in Setting~3 and 4 correspond to the extra continuous covariates.

Given the stochastic nature of the SAEM algorithm, its convergence is intrinsically related to the choice of coefficients $\gamma_q$, which must meet conditions that are not very demanding \citep{Kuhn2005}. In our implementation, we will define these coefficients as $\gamma_q=1$ if $q\leq K_1$ , and  $\gamma_q=\frac{1}{q-K_1}$ if $K_1<q\leq K_1+K_2$, where $K_1+K_2$ is the total number of iterations. With regard to the choice of $K_1$ and $K_2$, inspired by the existing implementation of SAEM in the \texttt{saemix} package \citep{Comets2017}, we will set $K_1=750$ and $K_2=250$. This allows for free fluctuation in the first iterations of the algorithm, forcing convergence when a point close to a (local) optimum has been reached.

For comparison, \texttt{glmmTMB} and \texttt{gamlss} are run using their default optimization settings; both converge rapidly for most designs, but they rely on deterministic optimization rather than the stochastic approximation scheme employed by SAEM, which makes SAEM more robust in small-sample or highly zero-inflated scenarios.

\subsection{Results}\label{sec:sim_results}

Figures~\ref{fig.31}--\ref{fig.32a} below (and Tables 1 and 2 in Appendix B of Supplementary Materials) summarize the finite-sample behavior of the proposed ZIBBMR--SAEM method (hereafter, SAEM), compared with \texttt{glmmTMB} and \texttt{gamlss}.

\begin{figure*}[htbp]%
\centering
\includegraphics[width=\linewidth]{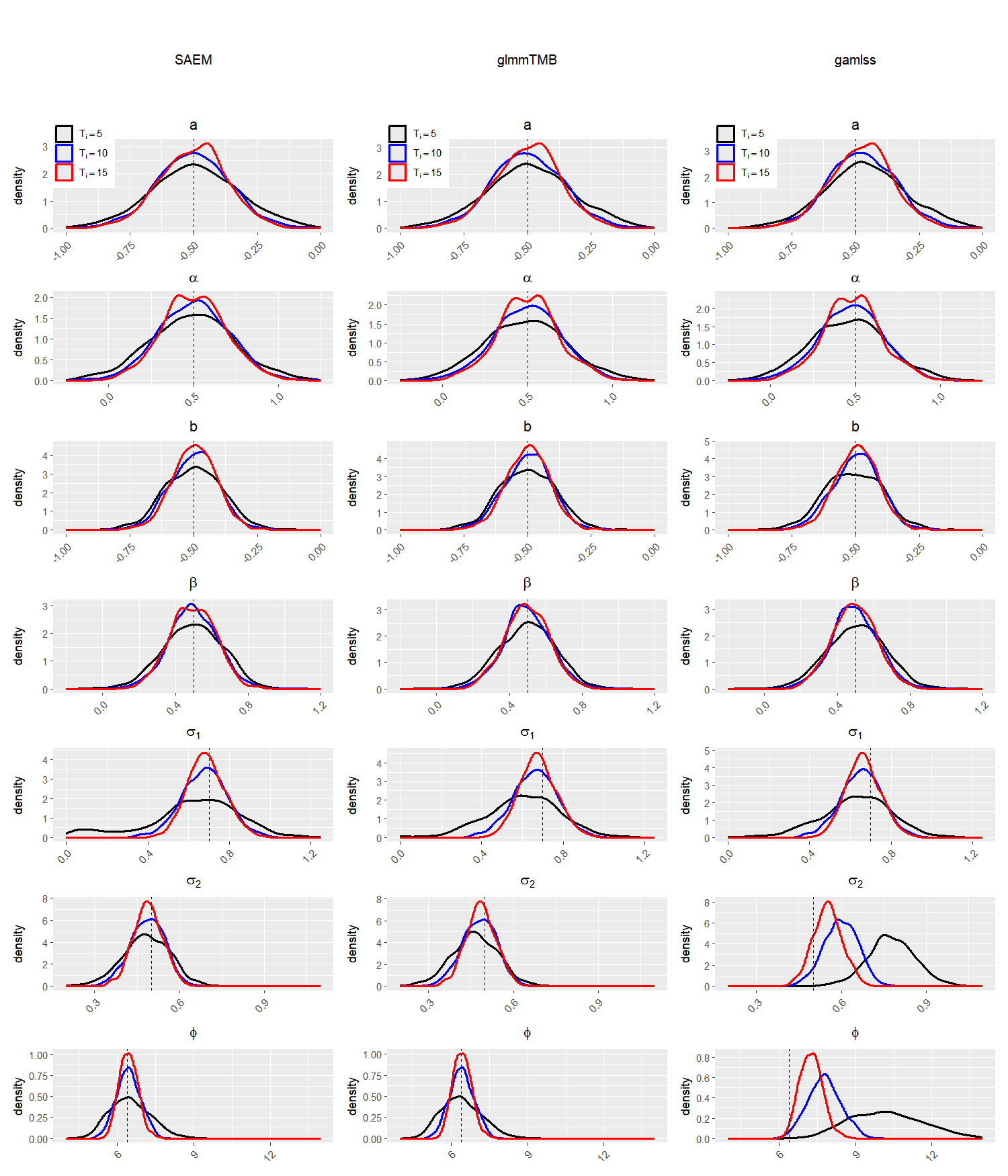}
\caption{Empirical densities of parameter estimates in Setting~1 for the proposed SAEM routine, \texttt{glmmTMB}, and \texttt{gamlss}.  
Parameters are grouped into: zero-inflation component ($a$, $\alpha$), mean component ($b$, $\beta$), and variance and dispersion component parameters ($\sigma_1^2$, $\sigma_2^2$, $\phi$).  
The vertical dotted lines indicate the true parameter values.}
\label{fig.31}
\end{figure*}

\begin{figure*}[htbp]%
\centering
\includegraphics[width=\linewidth]{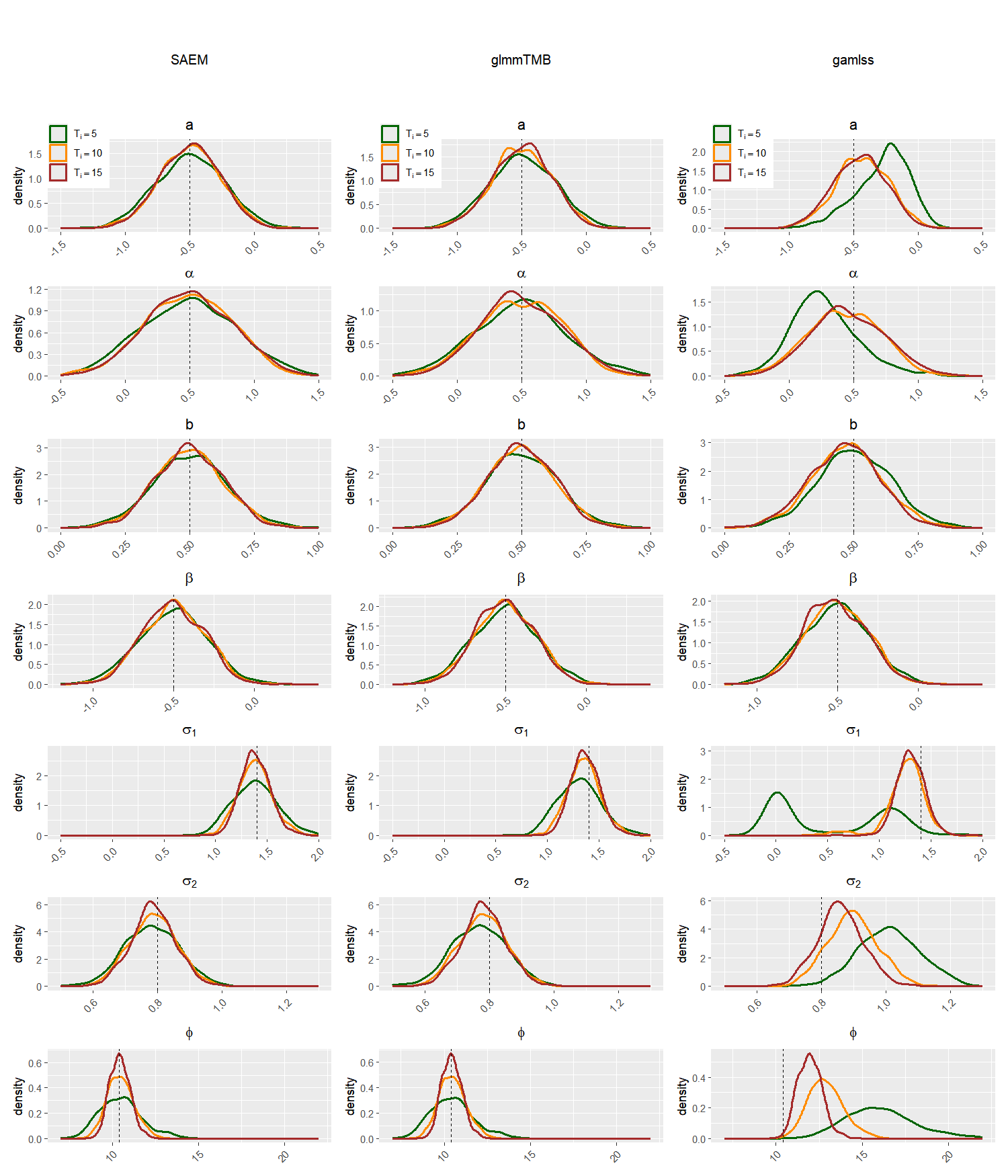}
\caption{Empirical densities of parameter estimates in Setting~2 for the proposed SAEM routine, \texttt{glmmTMB}, and \texttt{gamlss}.  
Grouping and interpretation of parameters follow Figure~\ref{fig.31}.}
\label{fig.32}
\end{figure*}

\begin{figure*}[htbp]%
\centering
\includegraphics[width=\linewidth]{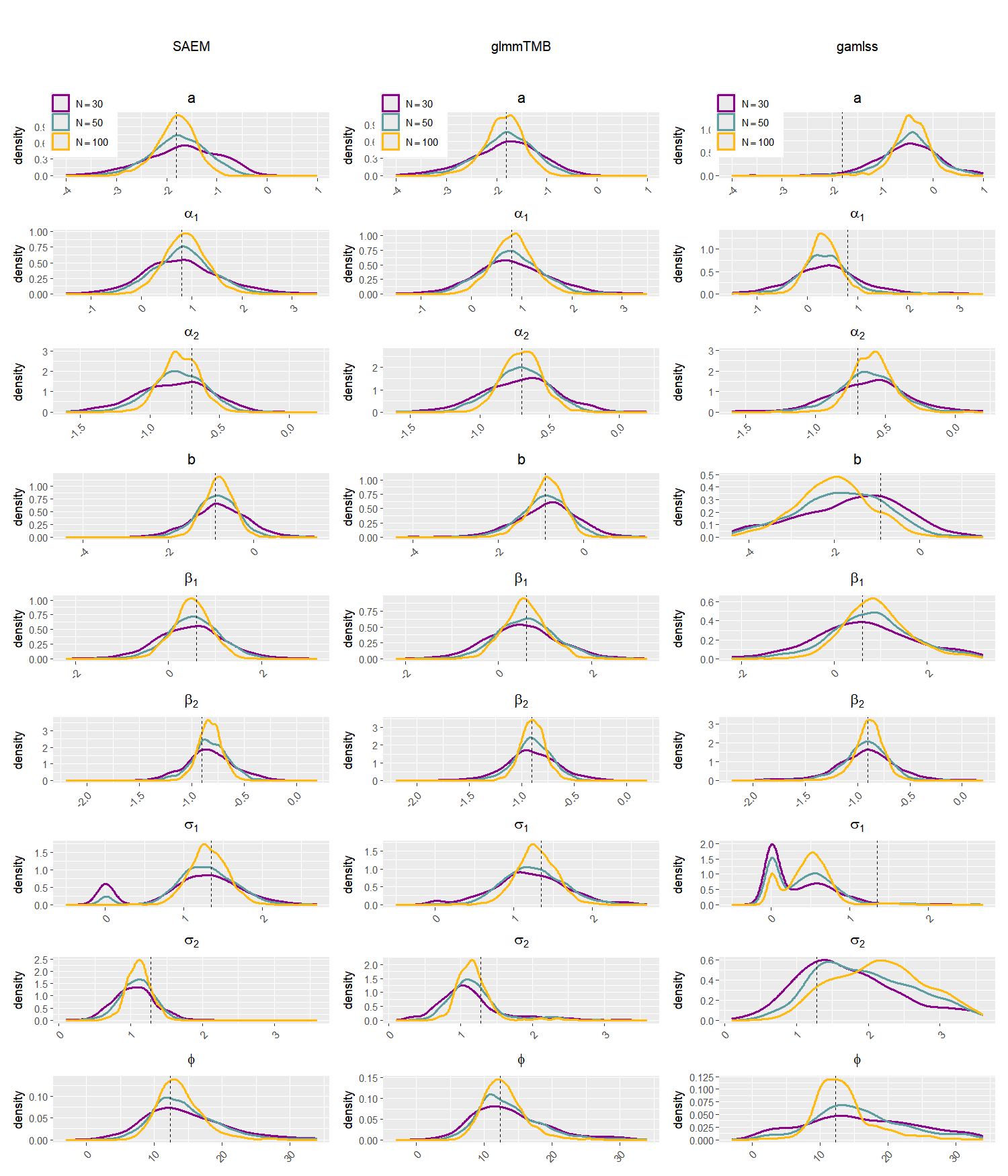}
\caption{Empirical densities of parameter estimates in Setting~3 for the proposed SAEM routine, \texttt{glmmTMB} and \texttt{gamlss}.  
Grouping and interpretation of parameters follow Figure~\ref{fig.31}.}
\label{fig.32a}
\end{figure*}

\paragraph{Zero-inflation component parameters}  
For the zero-inflation component intercept ($a$) and slopes ($\alpha$), all three methods (SAEM, \texttt{glmmTMB}, and \texttt{gamlss}) deliver broadly comparable performance across scenarios.  \texttt{gamlss} often achieves the smallest RMSE and mean absolute error (MAE) in moderate- to large-$T$ settings, whereas SAEM tends to produce the smallest bias, notably when the number of repeated measurements per subject is small.  

\paragraph{Mean component parameters}  
For the mean component intercept ($b$) and slopes ($\beta$), \texttt{glmmTMB} frequently achieves the lowest bias and RMSE, 
particularly when $T$ is large.  In small-$T$ designs, however, SAEM matches or outperforms \texttt{glmmTMB} in both bias and RMSE, indicating greater stability in settings with limited longitudinal information. 

\paragraph{Dispersion and variance component parameters}  
For the dispersion parameter ($\phi$) and variance components ($\sigma_1^2$, $\sigma_2^2$), SAEM generally achieves the smallest bias and error measures across settings. \texttt{glmmTMB} performs competitively when $\phi$ is large, but \texttt{gamlss} shows clear bias in $\phi$ estimation in small-$T$ or low-dispersion designs.  

\paragraph{Log-likelihood calculation} 
As mentioned above, we used Setting 3 not only to verify the estimation under continuous covariates, but also to compare the log-likelihood estimates provided by SAEM and \texttt{glmmTMB}. The use of penalized likelihood prevents considering \texttt{gamlss} in this comparison. As can be consulted in Table \ref{tab.33}, both methods obtained similar results in this regard in this simulation scenario. However, it should be noted that \texttt{glmmTMB} failed to provide values for the log-likelihood in up to 10\% of the datasets used, while SAEM, by using a robust importance sampling method and not requiring numerical approximations for this value, always provides an estimate of the log-likelihood of the model.

\paragraph{Type~I error assessment}  
Under Setting~4 (Table~\ref{tab.33}), both SAEM and \texttt{glmmTMB} yield empirical Type~I error rates close to nominal levels for Wald tests of single-parameter hypotheses and for likelihood ratio tests (LRT) of joint hypotheses.  
The LRT is not available for \texttt{gamlss} due to its penalized likelihood framework. In contrast, \texttt{gamlss} shows inflated Type~I error rates for Wald tests, likely due to bias introduced by penalization.

\begin{table}[hbtp]
\small
\centering
\caption{Empirical Type~I error rates for testing $H_0:\alpha_1=0$, $H_0:\beta_1=0$, and $H_0:\alpha_1=\beta_1=0$ under Setting~3 using SAEM, \texttt{glmmTMB}, and \texttt{gamlss}. Wald tests are used for the first two hypotheses; the likelihood ratio test (LRT) is used for the joint null (not available for \texttt{gamlss}). Nominal levels are 0.05 and 0.01.}
\label{tab.33}
\begin{tabular}{llcccccc}
\toprule
&&\multicolumn{2}{c}{\textbf{SAEM}}&\multicolumn{2}{c}{\textbf{\texttt{glmmTMB}}}&\multicolumn{2}{c}{\textbf{\texttt{gamlss}}}\\
\cmidrule{3-8}
&&\multicolumn{2}{c}{\textbf{Significance level}}&\multicolumn{2}{c}{\textbf{Significance level}}&\multicolumn{2}{c}{\textbf{Significance level}}\\
Null hypotheses && $0.05$ & $0.01$ & $0.05$ & $0.01$& $0.05$ & $0.01$\\
\midrule
 $H_0:\alpha_1=0$& $N=30$ & 0.078 & 0.025 & 0.076 & 0.015 & 0.284 & 0.170 \\
                 & $N=50$ & 0.060 & 0.019 & 0.061 & 0.018 & 0.328 & 0.195 \\
 $H_0:\beta_1=0$& $N=30$ & 0.077 & 0.022 & 0.081 & 0.019 & 0.168 & 0.067 \\
                & $N=50$ & 0.080 & 0.022 & 0.085 & 0.023 & 0.159 & 0.058 \\
 $H_0:\alpha_1=\beta_1=0$& $N=30$ & 0.063 & 0.012 & 0.069 & 0.014 &&\\
                         & $N=50$ & 0.063 & 0.017 & 0.070 & 0.018 &&\\
\bottomrule
\end{tabular}
\end{table}

\paragraph{Average computation time} 
With regard to the execution times for each routine, the average was calculated based on 1000 executions of each technique for Setting 1, with the other settings behaving similarly. Given this, for $T_i=5\ (10, 15)$ observations per individual, SAEM took $13.29\ (23.14, 36.01)$ seconds on average, while \texttt{glmmTMB} took $1.05\ (1.56,2.03)$ seconds to complete the estimation, although not necessarily to reach convergence, as seen in the log-likelihood results. It is noteworthy that \texttt{gamlss} maintained uniform behavior for all values of $T_i$, taking $13.22\ (13.16,13.79)$ seconds, also on average.

Overall, these results show that while all methods perform adequately for the zero-inflation and mean components in well-powered designs, the SAEM-based estimation routine consistently provides the most reliable estimates for variance and dispersion parameters, and offers tangible computational advantages in challenging scenarios.

\section{Application to Microbiome Count Data}\label{sec:realdata}

\subsection{Background and motivation}\label{sec:realdata_background}

The human vaginal microbiota is a complex ecosystem that plays a crucial role in women's reproductive health. Its composition and stability during pregnancy are of particular significance, as they can substantially influence perinatal outcomes, including preterm birth and infections. The advent of 16S rRNA gene sequencing has enabled detailed characterization of bacterial diversity and temporal dynamics, overcoming the limitations of culture-based methods.

In the study in \cite{Romero2014}, the vaginal microbiota of 22 healthy pregnant women who underwent full-term deliveries was longitudinally compared with that of 32 non-pregnant women. Using high-throughput sequencing and statistical models for longitudinal count data, it was found that the vaginal microbiota during pregnancy is characterized by greater stability and higher abundance of \textit{Lactobacillus} species (\textit{L. crispatus}, \textit{L. gasseri}, \textit{L. vaginalis}, and \textit{L. jensenii}). In contrast, non-pregnant women displayed higher diversity and greater prevalence of taxa associated with dysbiosis, such as \textit{Gardnerella}, \textit{Atopobium}, and \textit{Prevotella}.

This work was among the first to analyze the longitudinal vaginal microbiome in pregnant women, whereas most earlier studies focused on young, non-pregnant women \citep{Srinivasan2010,Ravel2011} and relied on culture-based approaches. The study has been widely cited for its methodological novelty and conclusions \citep{Florova2021,Kroon2018,Zhu2022}, and its original data remain publicly available \citep{Yi2024}.

\begin{table}[ht]
\caption{Description of the pregnant and non-pregnant women from the study of \cite{Romero2014}.}
\label{tab.31}
\centering
\begin{threeparttable}
\begin{tabular*}{0.8\linewidth}{@{\extracolsep{\fill}}lcc}
\toprule
 & \textbf{Non-pregnant} & \textbf{Pregnant} \\
\textbf{Characteristic}\textsuperscript{\textit{1}} & N = 32 & N = 22 \\ 
\midrule\addlinespace[2.5pt]
Age & 37 (31--43) & 24 (20--29) \\ 
N. of observations & 24 (21--29) & 6 (6--7) \\
Ethnic origin &  &  \\ 
\hspace*{1cm}Black & 16 (50\%) & 19 (86\%) \\ 
\hspace*{1cm}Hispanic, others & 3 (9.4\%) & 0 (0\%) \\ 
\hspace*{1cm}White & 13 (41\%) & 3 (14\%) \\ 
Nugent score\textsuperscript{\textit{2}} &  &  \\ 
\hspace*{1cm}$<7$ & 16 (50\%) & 19 (86\%) \\ 
\hspace*{1cm}$\geq7$ & 16 (50\%) & 3 (14\%) \\ 
\bottomrule
\end{tabular*}

\begin{tablenotes}\scriptsize
\item \textsuperscript{\textit{1}} Age: mean (Q1--Q3); ethnic group and Nugent score: n (\%).  
\item \textsuperscript{\textit{2}} At least one sample with Nugent score $\geq 7$ (bacterial vaginosis diagnosis \citep{Nugent1991}).
\end{tablenotes}
\end{threeparttable}
\end{table}


Despite its strengths, the study has limitations. As shown in Table~\ref{tab.31}, most participants were of African American origin, a factor linked to microbiota composition through genetic and environmental influences \citep{Serrano2019}. The small sample size also limits generalizability. In addition, the original mixed models included pregnancy as the sole covariate, omitting other potentially relevant factors such as age. For taxa with many zeros, zero-inflated models were used but without random effects for the zero component.

Later, Zhang et al.\citep{Zhang2020a} applied several mixed model specifications to these data, including linear and negative binomial models without zero inflation, and a Gaussian zero-inflated model for log-transformed abundances. However, this transformation is discouraged for count data \citep{McMurdie2014}, and the analysis focused only on abundance effects (via Wald tests), ignoring potential covariate influences on zero inflation.


\subsection{Microbiome model fitting and comparison}\label{sec:realdata_fit}
 
\begin{figure*}[hbtp]
    \centering
    \begin{subfigure}[t]{0.48\textwidth}
        \centering
        \includegraphics[width=\textwidth]{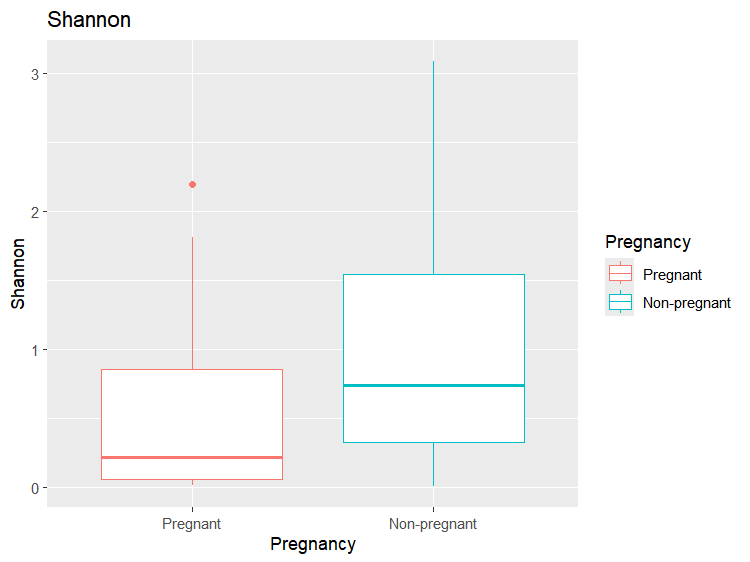}
        \caption{Shannon}
    \end{subfigure}%
    ~ 
    \begin{subfigure}[t]{0.48\textwidth}
        \centering
        \includegraphics[width=\textwidth]{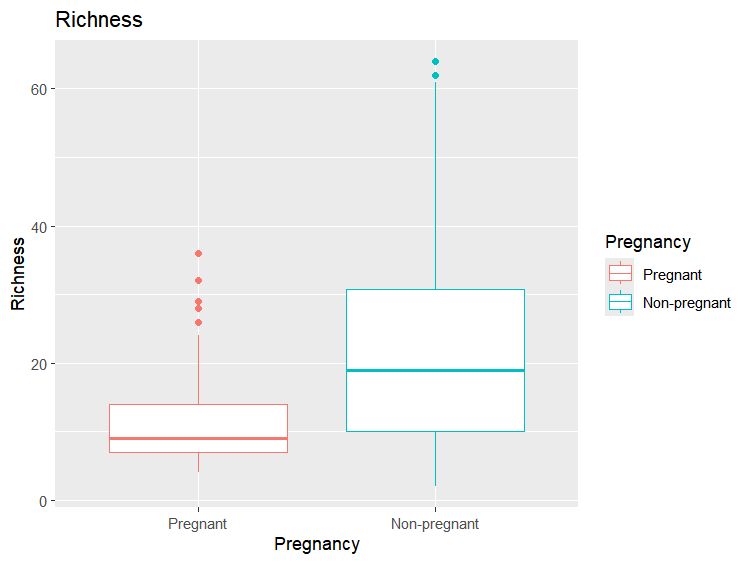}
        \caption{Richness}
    \end{subfigure}
    \\
    \begin{subfigure}[t]{0.48\textwidth}
        \centering
        \includegraphics[width=\textwidth]{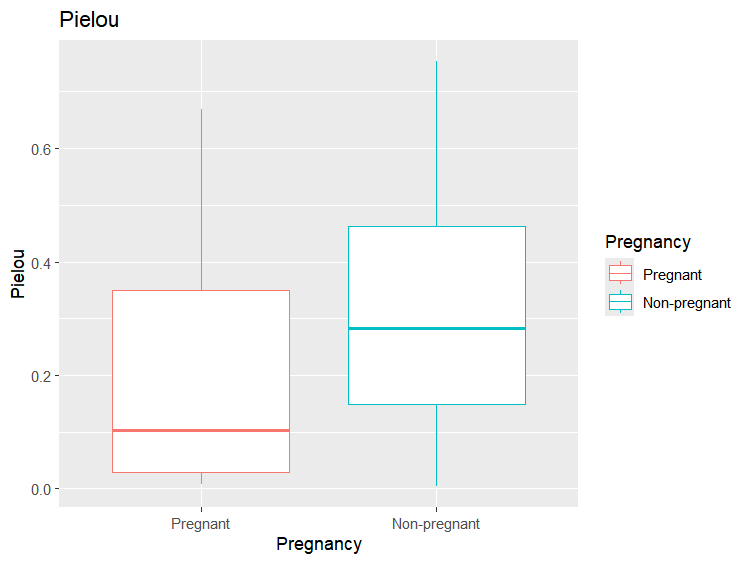}
        \caption{Pielou}
    \end{subfigure}%
    ~ 
    \begin{subfigure}[t]{0.48\textwidth}
        \centering
        \includegraphics[width=\textwidth]{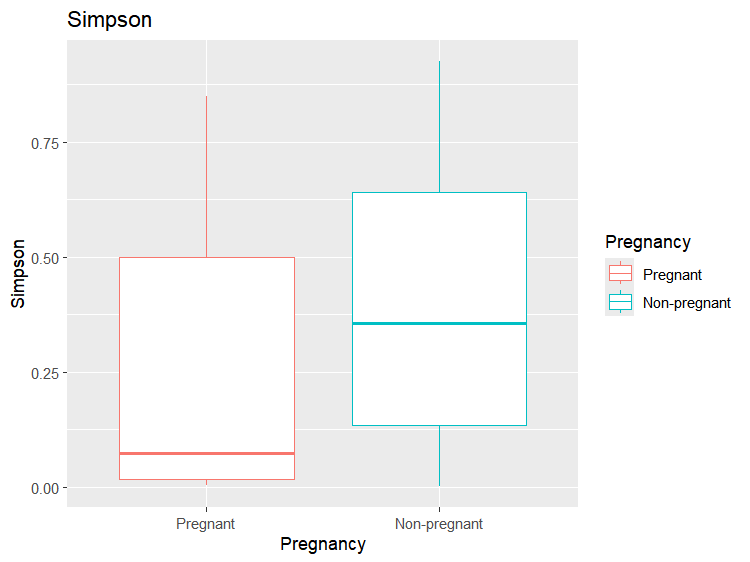}
        \caption{Simpson}
    \end{subfigure}
    \caption{Alpha-diversity indices for pregnant and non-pregnant women.}\label{fig.33}
\end{figure*}

We analyze the dataset from \cite{Romero2014}. Since Table \ref{tab.31} presented some of the important covariate characteristics in this dataset, Figure \ref{fig.33} now shows alpha-diversity indices for observations separated by pregnant and non-pregnant women. As is well known, alpha-diversity is an ecological measure that quantifies species diversity within a single sample or biological community, and can be expressed using indices that favor more diverse communities (Richness, Chao1) as well as those that focus on more balanced communities (Shannon, Simpson) \citep{Thukral2017}. As can be seen in this figure, all indices show greater diversity in non-pregnant women in terms of both parity and species richness. This is consistent with \cite{Ravel2011}, which mentions that stability in the vaginal microbiota is a common characteristic among pregnant women, regardless of their ethnic origin.

We consider the following specifications for the study:
\begin{itemize}
    \item \textbf{Model 1:} pregnancy, time and age as covariates, taking pregnancy as a factor of interest for testing.
    \item \textbf{Model 2:} pregnancy, time, age and interaction between time and pregnancy as covariates, testing the effect of pregnancy and the interaction.
\end{itemize}

Due to the nature of the models considered, only taxa with a proportion of zeros between 0.1 and 0.9 were selected, and any bacteria present in only one of the two groups of women were discarded. Thus, 57 taxa were selected in which the models specified above will be used. Initially, we considered using the ZIBBMR estimate using \texttt{glmmTMB} as a comparison (\texttt{gamlss} was discarded because it did not approximate the same log-likelihood function); however, this estimate gave inconsistent results, resulting in failures in the calculation of the log-likelihood in 54 of the 57 taxa considered. Therefore, we will use another model used for microbiota data, the Zero--Inflated Beta Regression (ZIBR) model \citep{Chen2016,Barrera2025}, which models zero inflation and the conditional mean on proportion data with a Beta distribution for the non-zero component. ZIBR is included solely as an external comparator; it differs from the proposed ZIBBMR in both data input (proportions vs.\ counts) and distributional assumptions (Beta vs.\ Beta--Binomial).

Once the respective LRTs were developed with a significance value of 0.05, SAEM-ZIBR detected 29 taxa in Model 1 in which pregnancy is significant, the same number as SAEM-ZIBBMR. For Model 2, SAEM-ZIBR detected nine and 15 taxa for pregnancy and interaction, respectively, while SAEM-ZIBBMR detected nine and 11 taxa, respectively. While the results appear similar, it should be noted that the bacterial species detected in both models are not necessarily the same, as summarized in Table \ref{tab.34}.

\begin{table}[hbtp]
\small
\centering
\caption{Bacterial taxa detected exclusively by ZIBR or ZIBBMR using the SAEM algorithm, for the covariates of interest in the two specifications considered.}\label{tab.34}
\begin{tabular}{cccccc}
\toprule
\multicolumn{3}{c}{\textbf{Only detected by ZIBBMR}}&\multicolumn{3}{c}{\textbf{Only detected by ZIBR}}\vspace*{1mm}\\
\midrule
\textbf{Model 1}&\multicolumn{2}{c}{\textbf{Model 2}}&\textbf{Model 1}&\multicolumn{2}{c}{\textbf{Model 2}}\\
\cmidrule{2-3}\cmidrule{5-6}
Pregnancy&Pregnancy&Interaction&Pregnancy&Pregnancy&Interaction\\
\midrule
\textit{Lachnospiraceae}&\textit{Lachnospiraceae}&\textit{Prevotella genogroup 3}&\textit{Aerococcus}&\textit{Aerococcus}&\textit{Aerococcus}\\
\textit{Lactobacillus vaginalis}&\textit{Staphylococcus}&\textit{Sneathia sanguinegens}&\textit{Bacteroidales}&\textit{Lactobacillus}&\textit{Clostridiales}\\
&&&&&\textit{Finegoldia magna}\\
&&&&&\textit{Peptoniphilus}\\
&&&&&\textit{Atopobium}\\
&&&&&\textit{Proteobacteria}\\
\bottomrule
\end{tabular}
\end{table}

The results indicate that the majority of the detected bacteria are the same for both methods and specifications. Therefore, it is worth analyzing the bacteria unique to each proposed model. For ZIBR, the most notable unique bacterium is \textit{Aerococcus}, which was detected in all three covariates of interest. Wang et al.~\cite{Wang2020} mentions that, when \textit{Lactobacillus} increases in abundance in the vaginal microbiota for various reasons (e.g., aerobic vaginitis or pregnancy), \textit{Aerococcus} decreases as a compensatory response. However, Huang et al.~\cite{Huang2023} disputes this claim, pointing out that the evidence of pregnancy's influence on \textit{Aerococcus} is limited to the data from Romero, which we examined here, and is not significant in other similar studies.

The most notable taxon detected by ZIBBMR is \textit{Lachnospiraceae}. A study \cite{Yang2023} compared the gut microbiota of women with preterm and normal births, and found that women with normal births had higher abundances of \textit{Lachnospiraceae}, showing a notable correlation. According to \cite{Sorbara2020}, this may result from their role as producers of anti-inflammatory metabolites in the human body. However, this role is fulfilled by a different microbial colony than the one we are analyzing here, but this is a finding that deserves our attention. ZIBBMR also detects \textit{Prevotella genogroup 3} and \textit{Sneathia sanguinegens}.

\section{Conclusions}\label{sec36}

We introduced a Zero--Inflated Beta--Binomial Mixed Regression (ZIBBMR) model tailored for longitudinal microbiome count data and proposed an estimation and inference strategy based on the SAEM algorithm. The model accounts for both zero inflation and overdispersion while incorporating subject-specific random effects in the zero-inflation and mean components. Simulation studies demonstrated that the proposed method provides accurate parameter estimates, competitive inference performance, and robustness in small-sample settings, with notable advantages over alternative implementations such as \texttt{glmmTMB} and \texttt{gamlss}. 

Applying the ZIBBMR model to a longitudinal human vaginal microbiome dataset identified taxa whose abundance and zero-inflation patterns were associated with pregnancy status, after adjusting for gestational age and other demographic covariates. Comparison with the external ZIBR benchmark \citep{Barrera2025} revealed overlapping yet distinct sets of significant taxa, reflecting the complementary strengths of count-based (ZIBBMR) and proportion-based (ZIBR) approaches. Given their different data inputs and distributional assumptions, applying both models in parallel can enhance the robustness of inference in microbiome studies when both counts and relative abundances are available.

The method also proved more reliable than \texttt{glmmTMB}, which failed to produce valid likelihood or standard error values for most taxa in this dataset. These results underline the suitability of the proposed approach for microbiome studies, particularly in contexts where high zero inflation and overdispersion are coupled with repeated measures.

A notable strength of the ZIBBMR framework is its direct use of raw sequencing counts, avoiding the statistical limitations associated with proportion-based transformations, which are prone to spurious correlations \citep{McMurdie2014,Gloor2016}. By leveraging the beta--binomial distribution, the model naturally accommodates the overdispersed and correlated structure characteristic of microbiome count data, and it offers a flexible foundation for future methodological extensions.

Several promising directions merit further exploration. From a modeling perspective, incorporating correlated random slopes or cross-taxon correlation structures could capture ecological dependencies between taxa. Extending the framework to handle multivariate responses or to integrate phylogenetic information would enhance both biological interpretability and generalizability. On the computational side, developing adaptive SAEM schemes or hybrid EM--variational approaches may improve scalability and convergence speed for large-scale, high-throughput sequencing datasets. From an applied standpoint, evaluating the joint use of ZIBBMR and ZIBR as complementary tools could yield more robust taxon detection and richer biological insights. Finally, beyond microbiome research, the proposed framework could be applied to other domains where longitudinal count data with zero inflation are prevalent, such as epidemiology and ecology.

\section*{Software}
An \textsf{R} implementation of the ZIBBMR model and the associated SAEM estimation procedure, including functions for simulation, model fitting, inference, and diagnostics, is available at \url{https://github.com/jbarrera232/saem-zibbmr}. The code reproduces the data analysis results presented in Section \ref{sec:realdata} of this paper and includes example scripts for fitting both ZIBBMR and ZIBR models to microbiome datasets. The package depends on standard \textsf{R} libraries for mixed-effects modeling and numerical optimization, and is designed to be easily extensible for related zero-inflated mixed model specifications.

\section*{Supplementary material}

Supplementary material associated with this article include derivations for the augmented case of the model, and additional tables for Section 3.

\section*{Funding}
Ana Arribas-Gil gratefully acknowledges the support of grant PID2021-123592OB-I00 funded by MCIN/AEI/10.13039
/501100011033 and by ERDF - A way of making Europe. Dae-Jin Lee research is supported by the Spanish State Research Agency (AEI) under the project SPHERES with project number PID2023-153222OB-I00. John Barerra and Cristian Meza received support from the ANID MATH-AmSud AMSUD 230032-SMILE project.


\end{document}


\maketitle
\vspace*{3cm}

\begin{appendices}

\section{Hierarchical augmentation of the Simulation step of SAEM}\label{secap1}
An alternative formulation of the simulation step leverages the hierarchical specification in Eq. (7) by augmenting the unobserved data with the latent variables
\[
\mathbf{w} = \{ w_{it} : 1 \le i \le N, \; 1 \le t \le T_i \},
\]
where $w_{it}$ denotes the latent Binomial success probability.  
Under this augmentation, the complete-data likelihood can be written as
\begin{equation}\label{equ.decomp3alt}
\begin{split}
p(\mathbf{Y}, \mathbf{w}, \boldsymbol{\varphi}; \theta)
&= p(\mathbf{Y} \mid \mathbf{w}, \boldsymbol{\varphi}; \alpha) \;
   p(\mathbf{w} \mid \boldsymbol{\varphi}; \beta, \phi) \;
   p(\boldsymbol{\varphi} \mid \boldsymbol{\mu}, \mathbf{G}) \\
&\propto \prod_{i,t} (1-p_{it})^{\mathbbm{1}_{\{Y_{it}=0\}}} \;
                  p_{it}^{\mathbbm{1}_{\{Y_{it}>0\}}} \;
                  f_1(Y_{it}; S_{it}, w_{it})^{\mathbbm{1}_{\{Y_{it} > 0\}}} \\
&\quad \times \prod_{i,t} f_2(w_{it}; u_{it}, \phi)^{\mathbbm{1}_{\{Y_{it} > 0\}}} \\
&\quad \times |\mathbf{G}|^{-N/2} \prod_{i} 
    \exp\!\left\{ -\frac{1}{2} (\varphi_i - \boldsymbol{\mu})^\top \mathbf{G}^{-1} (\varphi_i - \boldsymbol{\mu}) \right\},
\end{split}
\end{equation}
where $f_1$ is the Binomial pmf and $f_2$ is the Beta pdf from Eq. (6).

\paragraph{Gibbs-type simulation}
Following \citep{Meza2007}, the augmented SAEM simulation step proceeds as follows:
\begin{enumerate}
\item Sample $\boldsymbol{\varphi}^{(q+1)} \sim p(\cdot \mid \mathbf{Y}, \mathbf{w}^{(q)}; \theta^{(q)})$ using a Metropolis--Hastings algorithm.
\item Sample $\mathbf{w}^{(q+1)} \sim p(\cdot \mid \mathbf{Y}, \boldsymbol{\varphi}^{(q+1)}; \theta^{(q)})$, which admits the closed-form Beta conditional distribution:
\[
w_{it} \mid \mathbf{Y}, \boldsymbol{\varphi} \;\sim\; \mathrm{Beta}\!\left(Y_{it} + u_{it}\phi,\; S_{it} - Y_{it} + (1 - u_{it})\phi\right).
\]
\end{enumerate}

\paragraph{Metropolis--Hastings kernels}
For sampling $\boldsymbol{\varphi}$, three proposal kernels are employed and cycled for $m_1$, $m_2$, and $m_3$ iterations, respectively, at each SAEM iteration. This strategy balances global exploration with local refinement:
\begin{enumerate}
\item Prior proposal ($kern_1$): $N(\boldsymbol{\mu}^{(q)}, \mathbf{G}^{(q)})$.
\item Multivariate random walk ($kern_2$): $N(\varphi_{i,p-1}, \Omega^{(q)})$ with $\Omega^{(q)}$ tuned for optimal acceptance.
\item Univariate random walk ($kern_3$): perturb one randomly chosen component of $\varphi_{i,p-1}$ by $N(0,1)$ noise.
\end{enumerate}

\paragraph{Acceptance ratios}
For the \emph{original} specification without augmentation by $\mathbf{w}$, the Metropolis--Hastings acceptance ratio is given by:
\begin{align*}
\Delta_{i,p}^{(1)} 
&= \sum_{t=1}^{T_i} 
    \mathbbm{1}_{\{Y_{it}>0\}}
    \Bigg[
      \log\frac{
        B\!\left(Y_{it}+\hat{u}_{it}\phi^{(q)},\;
                S_{it}-Y_{it}+(1-\hat{u}_{it})\phi^{(q)}\right)
      }{
        B\!\left(Y_{it}+u_{it}\phi^{(q)},\;
                S_{it}-Y_{it}+(1-u_{it})\phi^{(q)}\right)
      } \\
&\hphantom{=\sum_{t=1}^{T_i} \mathbbm{1}_{\{Y_{it}>0\}} \Bigg[}
      - \log\frac{
        B\!\left(\hat{u}_{it}\phi^{(q)},\;
                (1-\hat{u}_{it})\phi^{(q)}\right)
      }{
        B\!\left(u_{it}\phi^{(q)},\;
                (1-u_{it})\phi^{(q)}\right)
      }
    \Bigg] \\
&\quad + \mathbbm{1}_{\{Y_{it}>0\}} \log\frac{\hat{p}_{it}}{p_{it}}
       + \mathbbm{1}_{\{Y_{it}=0\}} \log\frac{1-\hat{p}_{it}}{1-p_{it}} .
\end{align*}

with $\Delta_{i,p}^{(2)}$ for symmetric kernels obtained by adding the Gaussian prior correction.

For the \emph{augmented Gibbs} specification with $\mathbf{w}$:
\begin{align*}
\Delta_{i,p}^{(1)} &= \sum_{t=1}^{T_i} \mathbbm{1}_{\{Y_{it}>0\}} \left[
   \log\frac{\Gamma(u_{it}\phi^{(q)})\Gamma((1-u_{it})\phi^{(q)})\hat{p}_{it}}
             {\Gamma(\hat{u}_{it}\phi^{(q)})\Gamma((1-\hat{u}_{it})\phi^{(q)})p_{it}}
 + \phi^{(q)} (\hat{u}_{it} - u_{it}) \log\frac{w_{it}^{(q)}}{1-w_{it}^{(q)}} \right] \\
&\quad + \mathbbm{1}_{\{Y_{it}=0\}} \log\frac{1-\hat{p}_{it}}{1-p_{it}},
\end{align*}
with $\Delta_{i,p}^{(2)}$ analogous to the original case.

\section{Simulation study: Additional tables}

Complementing the results presented in section 3.7, the Tables~\ref{tab.32}--\ref{tab.32a} summarize the finite-sample behavior of the proposed ZIBBMR--SAEM method, compared with \texttt{glmmTMB} and \texttt{gamlss}, using the proposed ZIBBMR--SAEM method.

\begin{table}[htbp]
\scriptsize
\caption{Summary statistics of the results obtained by the SAEM algorithm and the packages \texttt{glmmTMB} and \texttt{gamlss} over 1000 simulation runs. For each parameter value and number of observations per individual, $T_i$, bold numbers indicate the lowest absolute value for each of bias, RMSE and MAE.}\label{tab.32}
\begin{tabular}{lclccccccccc}
\toprule
\textbf{Parameter} & \textbf{Value} & & \textbf{Bias} & \textbf{RMSE} & \textbf{MAE} & \textbf{Bias} & \textbf{RMSE} & \textbf{MAE} & \textbf{Bias} & \textbf{RMSE} & \textbf{MAE}\\
\toprule
&&&\multicolumn{3}{c}{\textbf{SAEM}}&\multicolumn{3}{c}{\texttt{glmmTMB}}&\multicolumn{3}{c}{\texttt{gamlss}}\vspace*{-1mm}\\
\cmidrule{4-6}\cmidrule{7-9}\cmidrule{10-12}
 $a$ & -0.5 & $T_i=5$ & \textbf{0.0076} & 0.1743 & 0.1381 & 0.0093 & 0.1700 & 0.1351 & 0.0394 & \textbf{0.1644} & \textbf{0.1301}\vspace*{-0.1mm}\\
 & & $T_i=10$ & \textbf{-0.0001} & 0.1419 & 0.1129 & 0.0011 & 0.1389 & 0.1109 & 0.0272 & \textbf{0.1340} & \textbf{0.1066}\vspace*{-0.1mm}\\
 & & $T_i=15$ & -0.0009 & 0.1296 & 0.1039 & \textbf{0.0002} & 0.1263 & 0.1009 & 0.0223 & \textbf{0.1224} & \textbf{0.0984}\vspace*{-0.1mm}\\

$\alpha$ & 0.5 & $T_i=5$ & \textbf{-0.0093} & 0.2474 & 0.1968 & -0.0104 & 0.2389 & 0.1921 & -0.0399 & \textbf{0.2273} & \textbf{0.1825}\vspace*{-0.1mm}\\
 & & $T_i=10$ & 0.0024 & 0.2031 & 0.1628 & \textbf{0.0013} & 0.1953 & 0.1560 & -0.0251 & \textbf{0.1863} & \textbf{0.1483}\vspace*{-0.1mm}\\
 & & $T_i=15$ & 0.0049 & 0.1837 & 0.1480 & \textbf{0.0043} & 0.1752 & 0.1395 & -0.0181 & \textbf{0.1682} & \textbf{0.1337}\vspace*{-0.1mm}\\

$b$ & -0.5 & $T_i=5$ & 0.0005 & 0.1136 & 0.0910 & \textbf{-0.0003} & \textbf{0.1103} & \textbf{0.0888} & -0.0090 & 0.1176 & 0.0932\vspace*{-0.1mm}\\
 & & $T_i=10$ & 0.0012 & 0.0955 & 0.0758 & \textbf{0.0006} & 0.0939 & 0.0743 & 0.0033 & \textbf{0.0939} & \textbf{0.0743}\vspace*{-0.1mm}\\
 & & $T_i=15$ & 0.0026 & 0.0857 & 0.0681 & \textbf{0.0015} & 0.0844 & 0.0669 & 0.0053 & \textbf{0.0837} & \textbf{0.0664}\vspace*{-0.1mm}\\

$\beta$ & 0.5 & $T_i=5$ & -0.0027 & 0.1622 & 0.1303 & \textbf{-0.0019} & \textbf{0.1550} & \textbf{0.1238} & 0.0053 & 0.1607 & 0.1285\vspace*{-0.1mm}\\
 & & $T_i=10$ & -0.0012 & 0.1336 & 0.1054 & \textbf{-0.0009} & \textbf{0.1281} & \textbf{0.1011} & -0.0035 & 0.1287 & 0.1018\vspace*{-0.1mm}\\
 & & $T_i=15$ & -0.0044 & 0.1233 & 0.1002 & \textbf{-0.0036} & 0.1191 & 0.0958 & -0.0072 & \textbf{0.1188} & \textbf{0.0955}\vspace*{-0.1mm}\\

$\sigma_1$ & 0.7 & $T_i=5$ & -0.0771 & 0.2466 & 0.1828 & -0.0931 & 0.2046 & 0.1589 & \textbf{-0.0718} & \textbf{0.1895} & \textbf{0.1454}\vspace*{-0.1mm}\\
 & & $T_i=10$ & \textbf{-0.0210} & 0.1135 & 0.0896 & -0.0377 & 0.1147 & 0.0909 & -0.0457 & \textbf{0.1106} & \textbf{0.0877}\vspace*{-0.1mm}\\
 & & $T_i=15$ & \textbf{-0.0135} & 0.0928 & \textbf{0.0743} & -0.0235 & 0.0933 & 0.0749 & -0.0376 & \textbf{0.0925} & 0.0750\vspace*{-0.1mm}\\

$\sigma_2$ & 0.5 & $T_i=5$ & \textbf{-0.0199} & \textbf{0.0925} & \textbf{0.0706} & -0.0375 & 0.0933 & 0.0730 & 0.2804 & 0.2967 & 0.2804\vspace*{-0.1mm}\\
 & & $T_i=10$ & \textbf{-0.0136} & \textbf{0.0632} & \textbf{0.0500} & -0.0207 & 0.0647 & 0.0511 & 0.0987 & 0.1156 & 0.1016\vspace*{-0.1mm}\\
 & & $T_i=15$ & \textbf{-0.0113} & \textbf{0.0532} & \textbf{0.0426} & -0.0150 & 0.0545 & 0.0436 & 0.0518 & 0.0728 & 0.0598\vspace*{-0.1mm}\\

$\phi$ & 6.4 & $T_i=5$ & \textbf{0.1208} & 0.8169 & 0.6489 & 0.0258 & \textbf{0.7992} & \textbf{0.6377} & 3.8619 & 4.1377 & 3.8619\vspace*{-0.1mm}\\
 & & $T_i=10$ & 0.0431 & 0.4741 & 0.3795 & \textbf{0.0157} & \textbf{0.4731} & \textbf{0.3791} & 1.3843 & 1.5273 & 1.3877\vspace*{-0.1mm}\\
 & & $T_i=15$ & 0.0456 & 0.3700 & 0.2961 & \textbf{0.0309} & \textbf{0.3689} & \textbf{0.2957} & 0.8584 & 0.9680 & 0.8618\vspace*{0.5mm}\\

&&&\multicolumn{3}{c}{\textbf{SAEM}}&\multicolumn{3}{c}{\texttt{glmmTMB}}&\multicolumn{3}{c}{\texttt{gamlss}}\vspace*{-1mm}\\
\cmidrule{4-6}\cmidrule{7-9}\cmidrule{10-12}

$a$ & -0.5 & $T_i=5$ & \textbf{0.0056} & 0.2603 & \textbf{0.2094} & 0.0084 & \textbf{0.2548} & 0.2039 & 0.2250 & 0.3043 & 0.2653\vspace*{-0.1mm}\\
 & & $T_i=10$ & 0.0043 & 0.2331 & 0.1867 & \textbf{0.0034} & 0.2255 & 0.1819 & 0.0678 & \textbf{0.2137} & \textbf{0.1720}\vspace*{-0.1mm}\\
 & & $T_i=15$ & \textbf{0.0029} & 0.2252 & 0.1820 & 0.0031 & 0.2183 & 0.1766 & 0.0476 & \textbf{0.2054} & \textbf{0.1671}\vspace*{-0.1mm}\\

$\alpha$ & 0.5 & $T_i=5$ & \textbf{-0.0057} & 0.3674 & 0.2953 & -0.0073 & 0.3543 & \textbf{0.2802} & -0.2220 & \textbf{0.3408} & 0.2889\vspace*{-0.1mm}\\
 & & $T_i=10$ & -0.0100 & 0.3370 & 0.2706 & \textbf{-0.0063} & 0.3175 & 0.2592 & -0.0719 & \textbf{0.2910} & \textbf{0.2354}\vspace*{-0.1mm}\\
 & & $T_i=15$ & \textbf{0.0001} & 0.3299 & 0.2653 & 0.0010 & 0.3105 & 0.2503 & -0.0433 & \textbf{0.2865} & \textbf{0.2329}\vspace*{-0.1mm}\\

$b$ & 0.5 & $T_i=5$ & 0.0015 & 0.1466 & 0.1157 & \textbf{0.0010} & \textbf{0.1425} & \textbf{0.1127} & 0.0059 & 0.1457 & 0.1150\vspace*{-0.1mm}\\
 & & $T_i=10$ & \textbf{-0.0029} & 0.1358 & 0.1075 & -0.0031 & \textbf{0.1330} & \textbf{0.1050} & -0.0235 & 0.1403 & 0.1106\vspace*{-0.1mm}\\
 & & $T_i=15$ & 0.0008 & 0.1290 & 0.1028 & \textbf{-0.0007} & \textbf{0.1254} & \textbf{0.1005} & -0.0371 & 0.1402 & 0.1106\vspace*{-0.1mm}\\

$\beta$ & -0.5 & $T_i=5$ & \textbf{-0.0017} & 0.2084 & 0.1667 & -0.0024 & \textbf{0.1976} & \textbf{0.1583} & -0.0099 & 0.2038 & 0.1631\vspace*{-0.1mm}\\
 & & $T_i=10$ & 0.0020 & 0.1925 & 0.1539 & \textbf{0.0013} & \textbf{0.1847} & \textbf{0.1478} & -0.0051 & 0.1910 & 0.1537\vspace*{-0.1mm}\\
 & & $T_i=15$ & -0.0050 & 0.1851 & 0.1492 & \textbf{-0.0031} & \textbf{0.1754} & \textbf{0.1416} & -0.0127 & 0.1863 & 0.1505\vspace*{-0.1mm}\\

$\sigma_1$ & 1.4 & $T_i=5$ & \textbf{-0.0137} & \textbf{0.2173} & \textbf{0.1737} & -0.0895 & 0.2238 & 0.1807 & -0.8534 & 1.0247 & 0.8755\vspace*{-0.1mm}\\
 & & $T_i=10$ & \textbf{-0.0158} & \textbf{0.1601} & \textbf{0.1269} & -0.0467 & 0.1617 & 0.1289 & -0.1458 & 0.2412 & 0.1739\vspace*{-0.1mm}\\
 & & $T_i=15$ & \textbf{-0.0144} & \textbf{0.1425} & \textbf{0.1135} & -0.0333 & 0.1429 & 0.1147 & -0.0967 & 0.1637 & 0.1322\vspace*{-0.1mm}\\
 
$\sigma_2$ & 0.8 & $T_i=5$ & \textbf{-0.0182} & \textbf{0.0884} & \textbf{0.0704} & -0.0310 & 0.0919 & 0.0732 & 0.2122 & 0.2339 & 0.2133\vspace*{-0.1mm}\\
 & & $T_i=10$ & \textbf{-0.0122} & \textbf{0.0741} & \textbf{0.0594} & -0.0185 & 0.0753 & 0.0605 & 0.0980 & 0.1245 & 0.1046\vspace*{-0.1mm}\\
 & & $T_i=15$ & \textbf{-0.0122} & \textbf{0.0688} & \textbf{0.0544} & -0.0163 & 0.0699 & 0.0553 & 0.0599 & 0.0926 & 0.0751\vspace*{-0.1mm}\\

$\phi$ & 10.4 & $T_i=5$ & 0.1150 & \textbf{1.2357} & \textbf{0.9758} & \textbf{0.0891} & 1.2472 & 0.9846 & 5.7136 & 6.0776 & 5.7136\vspace*{-0.1mm}\\
 & & $T_i=10$ & 0.0933 & \textbf{0.7848} & \textbf{0.6240} & \textbf{0.0917} & 0.7878 & 0.6267 & 2.4983 & 2.6995 & 2.4997\vspace*{-0.1mm}\\
 & & $T_i=15$ & 0.0537 & \textbf{0.5907} & \textbf{0.4705} & \textbf{0.0516} & 0.5909 & 0.4708 & 1.5690 & 1.7177 & 1.5714\vspace*{0.5mm}\\
 
\bottomrule
\end{tabular}
\end{table}

\begin{table}[htbp]
\scriptsize
\caption{Summary statistics of the results obtained by the SAEM algorithm and and the packages \texttt{glmmTMB} and \texttt{gamlss} over 1000 simulation runs. For each parameter value and number of individuals, $N$, bold numbers indicate the lowest absolute value for each of bias, RMSE and MAE.}\label{tab.32a}
\begin{tabular}{lclccccccccc}
\toprule
\textbf{Parameter} & \textbf{Value} & & \textbf{Bias} & \textbf{RMSE} & \textbf{MAE} & \textbf{Bias} & \textbf{RMSE} & \textbf{MAE} & \textbf{Bias} & \textbf{RMSE} & \textbf{MAE}\\
\toprule
&&&\multicolumn{3}{c}{\textbf{SAEM}}&\multicolumn{3}{c}{\texttt{glmmTMB}}&\multicolumn{3}{c}{\texttt{gamlss}}\vspace*{-1mm}\\
\cmidrule{4-6}\cmidrule{7-9}\cmidrule{10-12}
$a$ & -1.80 & $N=30$ & 0.1160 & 0.7644 & 0.6135 & \textbf{-0.0405} & \textbf{0.7423} & \textbf{0.5641} & 2.8743 & 15.8214 & 3.9015 \vspace*{-0.1mm}\\
& & $N=50$ & 0.0459 & 0.5463 & 0.4365 & \textbf{-0.0087} & \textbf{0.5450} & \textbf{0.4303} & 2.3195 & 16.8357 & 2.3344 \vspace*{-0.1mm}\\
& & $N=100$ & \textbf{-0.0133} & \textbf{0.3729} & \textbf{0.2930} & -0.0469 & 0.3820 & 0.2998 & 1.5483 & 2.9205 & 1.5484 \vspace*{-0.1mm}\\

$\alpha_1$ & 0.80 & $N=30$ & \textbf{0.0121} & 0.7619 & 0.5952 & 0.0169 & \textbf{0.7410} & \textbf{0.5777} & 3.2631 & 71.7294 & 7.6583 \vspace*{-0.1mm}\\
& & $N=50$ & 0.0442 & 0.5778 & 0.4520 & \textbf{0.0024} & \textbf{0.5679} & \textbf{0.4481} & -1.5974 & 30.2205 & 2.1587 \vspace*{-0.1mm}\\
& & $N=100$ & 0.0763 & 0.4131 & 0.3277 & \textbf{0.0235} & \textbf{0.3931} & \textbf{0.3115} & -0.6225 & 2.6617 & 0.6918 \vspace*{-0.1mm}\\

$\alpha_2$ & -0.70 & $N=30$ & -0.1262 & 0.2910 & 0.2285 & \textbf{-0.0110} & \textbf{0.2760} & \textbf{0.2145} & 0.2487 & 5.5525 & 0.6081 \vspace*{-0.1mm}\\
& & $N=50$ & -0.1042 & 0.2259 & 0.1768 & \textbf{-0.0112} & \textbf{0.2034} & \textbf{0.1586} & 2.3606 & 71.9637 & 2.4620 \vspace*{-0.1mm}\\
& & $N=100$ & -0.0822 & 0.1564 & 0.1250 & \textbf{0.0024} & \textbf{0.1395} & \textbf{0.1112} & 0.1088 & 0.1835 & 0.1417 \vspace*{-0.1mm}\\

$b$ & -0.90 & $N=30$ & 0.1056 & \textbf{0.6766} & \textbf{0.5229} & \textbf{0.0517} & 0.7610 & 0.5797 & -0.9298 & 1.9406 & 1.3621 \vspace*{-0.1mm}\\
& & $N=50$ & 0.0453 & \textbf{0.4834} & \textbf{0.3813} & \textbf{-0.0145} & 0.5748 & 0.4438 & -1.1801 & 1.8631 & 1.3633 \vspace*{-0.1mm}\\
& & $N=100$ & 0.0870 & \textbf{0.3464} & \textbf{0.2765} & \textbf{0.0257} & 0.4004 & 0.3144 & -1.2317 & 1.6162 & 1.3007 \vspace*{-0.1mm}\\

$\beta_1$ & 0.60 & $N=30$ & -0.1100 & \textbf{0.7413} & \textbf{0.5794} & \textbf{-0.0611} & 0.8246 & 0.6292 & 0.3391 & 2.7814 & 1.1914 \vspace*{-0.1mm}\\
& & $N=50$ & -0.0461 & \textbf{0.5612} & \textbf{0.4432} & \textbf{0.0169} & 0.6545 & 0.5079 & 0.3506 & 1.3624 & 0.9027 \vspace*{-0.1mm}\\
& & $N=100$ & -0.0926 & \textbf{0.4106} & \textbf{0.3235} & \textbf{-0.0118} & 0.4563 & 0.3577 & 0.3600 & 0.8875 & 0.6322 \vspace*{-0.1mm}\\

$\beta_2$ & -0.90 & $N=30$ & 0.0796 & \textbf{0.2504} & \textbf{0.1914} & \textbf{0.0059} & 0.2613 & 0.1995 & -0.0382 & 0.5451 & 0.2462 \vspace*{-0.1mm}\\
& & $N=50$ & 0.0803 & 0.1880 & 0.1478 & \textbf{0.0078} & \textbf{0.1797} & \textbf{0.1386} & -0.0135 & 0.2269 & 0.1647 \vspace*{-0.1mm}\\
& & $N=100$ & 0.0761 & 0.1324 & 0.1062 & \textbf{0.0074} & \textbf{0.1139} & \textbf{0.0902} & 0.0093 & 0.1384 & 0.1054 \vspace*{-0.1mm}\\

$\sigma_1$ & 1.35 & $N=30$ & -0.2125 & 0.6456 & 0.4726 & \textbf{-0.1283} & \textbf{0.5296} & \textbf{0.3991} & 3.2119 & 46.7426 & 5.1846 \vspace*{-0.1mm}\\
& & $N=50$ & -0.1032 & 0.4363 & 0.3201 & \textbf{-0.1000} & \textbf{0.3878} & \textbf{0.3113} & -0.3045 & 18.5523 & 1.5728 \vspace*{-0.1mm}\\
& & $N=100$ & \textbf{-0.0361} & \textbf{0.2359} & \textbf{0.1883} & -0.0626 & 0.2557 & 0.2024 & -0.8943 & 0.9369 & 0.8992 \vspace*{-0.1mm}\\

$\sigma_2$ & 1.28 & $N=30$ & -0.2987 & \textbf{0.5016} & \textbf{0.3763} & \textbf{-0.1585} & 0.5622 & 0.4284 & 1.1680 & 3.8981 & 1.3919 \vspace*{-0.1mm}\\
& & $N=50$ & -0.2019 & \textbf{0.3310} & \textbf{0.2547} & \textbf{-0.0951} & 0.3928 & 0.2939 & 1.5041 & 3.4468 & 1.5802 \vspace*{-0.1mm}\\
& & $N=100$ & -0.1671 & \textbf{0.2373} & \textbf{0.1967} & \textbf{-0.1009} & 0.2804 & 0.2083 & 1.6499 & 3.5445 & 1.6793 \vspace*{-0.1mm}\\

$\phi$ & 12.30 & $N=30$ & 4.4611 & 14.3424 & 6.8313 & \textbf{3.3669} & \textbf{12.6306} & \textbf{5.9997} & 7.2160 & 15.4420 & 10.2447 \vspace*{-0.1mm}\\
& & $N=50$ & 2.4394 & 5.9903 & 4.0989 & \textbf{1.5291} & \textbf{5.3352} & \textbf{3.6469} & 4.5668 & 9.4448 & 6.6060 \vspace*{-0.1mm}\\
& & $N=100$ & 1.4718 & 3.6310 & 2.6230 & \textbf{0.5794} & \textbf{3.1230} & \textbf{2.2986} & 1.3951 & 5.2315 & 3.3507 \vspace*{0.5mm}\\
\bottomrule
\end{tabular}
\end{table}

\end{appendices}
